\documentclass[preprint2]{aastex}

\usepackage{epsfig}

\shorttitle{}
\shortauthors{Haverkorn et al.}

\begin{document}

\title{The outer scale of turbulence in the magneto-ionized Galactic
 interstellar medium}

\author{M. Haverkorn\footnote{Marijke Haverkorn is a Jansky fellow of the National Radio Astronomy
  Observatory} \footnote{Astronomy Department University of California at
  Berkeley, 601 Campbell Hall, Berkeley CA 94720, USA,
  marijke@astro.berkeley.edu}, J. C. Brown\footnote{Centre for Radio
  Astronomy, University of Calgary, 2500 University Drive N.W.,
  Calgary, AB, Canada; jocat@ras.ucalgary.ca},
  B. M. Gaensler\footnote{School of Physics A29, The University of
  Sydney, NSW 2006, Australia}
  \footnote{Australian Research Council Federation Fellow},
  N. M. McClure-Griffiths\footnote{Australia Telescope National
  Facility, CSIRO, PO Box 76, Epping, NSW 1710, Australia;
  naomi.mcclure-griffiths@csiro.au}}

\begin{abstract}

We analyze Faraday rotation and depolarization of extragalactic radio
point sources in the direction of the inner Galactic plane to
determine the outer scale and amplitude of the rotation measure power
spectrum. Structure functions of rotation measure show lower
amplitudes than expected when extrapolating electron density
fluctuations to large scales assuming a Kolmogorov spectral index.
This implies an outer scale of those fluctuations on the order of a
parsec, much smaller than commonly assumed.  Analysis of partial
depolarization of point sources independently indicates a small outer
scale of a Kolmogorov power spectrum. In the Galaxy's spiral arms, no
rotation measure fluctuations on scales above a few parsecs are
measured. In the interarm regions fluctuations on larger scales than
in spiral arms are present, and show power law behavior with a shallow
spectrum. These results suggest that in the spiral arms stellar
sources such as stellar winds or protostellar outflows dominate the
energy injection for the turbulent energy cascade on parsec scales,
while in the interarm regions supernova and super bubble explosions
are the main sources of energy on scales on the order of 100~parsecs.

\end{abstract}


\section{Introduction}

Turbulence in the ionized phase of the interstellar medium (ISM) of
the Milky Way is well described on small scales, while its properties
on larger scales are more uncertain. On scales smaller than
$\sim10^{11}$~m ($\sim 10^{-5}$~pc), the turbulence in the ionized
medium is well characterized from diffractive and dispersive processes
in the ISM influencing pulsar signals. The comprehensive study by
\citet{ars95} showed that on scales of $10^5$ to $10^{13}$~m ($\sim
10^{-11}$~pc to $10^{-3}$~pc) the power spectrum of electron density
$n_e$ is well described by a power law with a spectral index
consistent with the Kolmogorov spectral index $\alpha=5/3$
\citep{k41}. Most observed electron density power spectra are
compatible with a Kolmogorov power spectrum
\citep[e.g.][]{sg90,ssh00,wmj05,yhc07}, although other spectral
indices have been reported \citep[e.g.][]{lkm01,sss03}.  On larger
scales the slope and extent of the electron density power spectrum are
much more uncertain. \citet{ars95} include measurements of \ion{H}{1}
and of Faraday rotation measures for which uncertain assumptions about
the correlation between \ion{H}{1} and $n_e$ and about magnetic
fields, respectively, needed to be made. Although these results
suggest a power law that connects to the Kolmogorov power spectrum on
small scales, these assumptions make the behavior of the power
spectrum at larger scales somewhat speculative.

Fluctuations in the magneto-ionized medium on parsec scales have been
measured using structure functions\footnote{A structure function
measures the amount of fluctuations in a quantity as a function of the
scale of the fluctuations. The second order structure function of a
function $f$ is defined as $D_f(\delta\theta) = \langle
(f(\theta)-f(\theta+\delta\theta))^2\rangle_{\theta}$, where $\theta$
is the position of a source in angular coordinates, $\delta\theta$ is
the separation between sources, i.e.\ the scale of the measured
fluctuation, and $\langle\rangle_{\theta}$ means the averaging over
all positions $\theta$.} of rotation measures (RMs) of extragalactic
sources. \citet{sc86} showed that the structure function of
high-latitude sources is flat, indicating that fluctuations in RM only
exist on size scales smaller than the scales they probe, viz.\ about
3$^{\circ}$.  This means that there is no contribution from
large-scale fluctuations from the Milky Way in the RM (although a
constant RM from the Milky Way is not ruled out), and that it is most
likely the RM contribution intrinsic to the sources that
dominates. This was confirmed by \citet{l87}. RM structure functions
at lower latitudes, however, show structure functions consistent with
a power law, although the slope tends to be shallower than a
Kolmogorov slope \citep{sc86,ccs92}. Other observations in or near the
Galactic plane \citep{sh04,hkb03,hgb06} also find shallow slopes for
RM structure functions.

Measurements of the outer scale of fluctuations, i.e.\ the scale at
which a structure function saturates, differ considerably across the
Galaxy.  \citet{ls90} used the autocorrelation function of synchrotron
radiation to derive a typical scale of 90~pc at a distance of 1~kpc in
a region near the north Galactic pole. \citet{hgb06} studied sources
in the Galactic plane and found that the outer scale of fluctuations
is smaller than about 10~pc for the spiral arms, while it is roughly
100~pc for the interarm regions.  A similar outer scale of $\sim90$~pc
in the magneto-ionized medium is found in the Large Magellanic Cloud
using Faraday rotation \citep{ghs05}, whereas \ion{H}{1} measurements
indicate a much larger outer scale of a few kpc in the Large
Magellanic Cloud \citep{eks01}, Small Magellanic Cloud \citep{ssd99},
and in external galaxies \citep{wca99,eel03}. \citet{hfm04} estimated
a magnetic energy spectrum with a slope of $-0.37$ up to scales of
15~kpc. They suggest that a magnetic energy spectrum which is flatter
than Kolmogorov on scales larger than the injection scale of
$10-100$~pc is dictated by magnetic helicity inversely cascading up
from the injection scale to larger scales. However, the pulsar
rotation measure and dispersion measure data they use for the power
law fit has a scatter of several orders of magnitude, making the
resulting spectrum uncertain.

Stellar sources of energy input are expected to dominate the turbulent
driving in the Milky Way, except in the outskirts of the Galaxy where
star formation is low and gravitational sources and instabilities such
as the magneto-rotational instability \citep{bh91,hb91} come into play
\citep{sb99}.  The stellar sources include supernovae, superbubbles,
stellar winds, protostellar outflows and \ion{H}{2} regions.
\citet{nf96} calculate a broadband source function mostly dominated by
supernovae, which is confirmed by \citet{mk04}.  The \citet{nf96}
source function shows a contribution of superbubbles to the turbulent
driving on scales above $\sim$~100~pc to about a kpc. Turbulent
driving on these scales is not observed, possibly because of the
finite extent of the Galaxy. So a maximum driving scale of about
100~pc due to supernovae is often implicitly assumed, although it is
not unlikely that these sources inject energy into the medium (also)
on smaller scales.

What, if any, is the relation of the electron density and magnetic
field fluctuations at these large scales to the Kolmogorov-like
spectrum below $10^{13}$~m?  Can both sets of observations be
reconciled with one spectrum from kilometer to parsec scales? What are
the characteristics of the fluctuations in the ionized ISM on larger
scales?

We address these questions in this paper, using RM structure functions
from extragalactic sources behind the inner Galactic plane. These data
are discussed in Section~\ref{s:data}. We determine the outer scale of
fluctuations using two independent methods: (1) the amplitude and
slope of the structure function indicate an uncommonly small outer
scale of Kolmogorov turbulence in the magneto-ionized ISM, as we will
explain in Section~\ref{s:sf}, and (2) the same conclusions can be
drawn from analysis of depolarization of extragalactic point sources
by the Galactic ISM, as shown in Section~\ref{s:depol}.
Section~\ref{s:steep} gives arguments for a steeper (Kolmogorov) power
spectrum on small scales, and a discussion of the results can be found
in Section~\ref{s:disc}. Section~\ref{s:sum} provides a summary and
conclusions.

\section{Data analysis of polarized extragalactic point sources}
\label{s:data}

The data used are from the Southern Galactic Plane Survey
\citep[SGPS,][]{mdg05,hgm06}, a neutral hydrogen and
full-po\-la\-ri\-za\-tion 1.4~GHz continuum survey of the Galactic
plane. The continuum part spans an area of $253^{\circ} < l <
357^{\circ}$ and $|b| < 1.5^{\circ}$ and contains 148 polarized
sources of which the rotation measure is measured unambiguously
\citep{bhg07}. The data were obtained with the Australia Telescope
Compact Array (ATCA) and are publicly
available\footnote{http://www.atnf.csiro.au/research/cont/sgps/queryForm.html}. For
more details on the data reduction see \citet{bhg07}.

Two corrections have been applied to our sample of extragalactic
sources.  Firstly, structure functions are sensitive to large-scale
gradients in electron density across the field of view and include a
geometrical component due to the change in direction of the regular
magnetic field, \citep[see e.g.\ ][]{bhg07}. As a first order
correction, we approximate this with a 2D linear gradient in RM, and
subtract this from the region over which a structure function is
computed. Furthermore, lines of sight through discrete structures like
H~{\sc ii} regions and supernova remnants can have deceptively large
RMs due to an increased electron density and possibly magnetic field
within these localized structures \citep{mwk03}. Therefore, we have
used the total intensity 1.4~GHz radio data from the ATCA combined
with Parkes single-dish data as well as H$\alpha$ maps \citep{f03} to
discard 27~extragalactic sources with a sight line passing through a
visible supernova remnant or H~{\sc ii} region. These sources are
listed in Table~\ref{t:flag}.  We recognize that omitting RMs through
discrete high density regions may introduce a bias in the structure
function, viz.\ decrease the structure function amplitude on large
scales. However, the results are very similar to results without
discarding extreme RMs \citep[cf.\ ][]{hgb06}, indicating that the
bias, if present, is low.

\begin{table*}[!t]
\begin{center}
\begin{tabular}{rrrl|rrrl|rrrl}
\multicolumn{1}{c}{$l$} & \multicolumn{1}{c}{$b$} & \multicolumn{1}{c}{RM} &  &
\multicolumn{1}{c}{$l$} & \multicolumn{1}{c}{$b$} & \multicolumn{1}{c}{RM} &  &
\multicolumn{1}{c}{$l$} & \multicolumn{1}{c}{$b$} & \multicolumn{1}{c}{RM} &  \\
\hline
355.42 &  -0.81 &   600.53 & H$\alpha$ & 308.93 &   0.40 &  -752.10 & I         &
 267.03 &   0.04 &   298.38 & I \\ 
351.82 &   0.17 &   134.43 & I         & 308.73 &   0.07 &  -661.47 & I         &
 263.22 &   1.08 &   826.49 & H$\alpha$ \\ 
337.19 &   0.02 &    56.25 & I         & 299.42 &  -0.23 &   534.50 & I         &
 263.20 &   1.07 &   739.48 & H$\alpha$ \\ 
337.06 &   0.85 &  -738.88 & I         & 295.29 &  -1.23 &   -43.36 & H$\alpha$ &
 260.69 &  -0.23 &   203.78 & H$\alpha$ \\ 
333.72 &  -0.27 &   204.08 & I         & 295.23 &  -1.05 &  -206.71 & H$\alpha$ &
 260.52 &  -0.55 &   247.36 & H$\alpha$ \\
332.14 &   1.03 &  -754.43 & I         & 294.38 &  -0.75 &   470.04 & H$\alpha$ &
 260.41 &  -0.43 &   221.22 & H$\alpha$ \\
329.48 &   0.22 &  -100.23 & I         & 294.29 &  -0.90 &   449.20 & H$\alpha$ &
 259.78 &   1.22 &   250.40 & H$\alpha$ \\
312.37 &  -0.03 &  -438.49 & I         & 288.27 &  -0.70 &   491.08 & H$\alpha$ &
 254.16 &  -0.34 &  -337.84 & H$\alpha$ \\
309.06 &   0.84 &  -504.17 & I         & 282.07 &  -0.78 &   861.69 & I         &
 253.68 &  -0.60 &  -348.90 & H$\alpha$ \\
\hline
\end{tabular}
\caption{Extragalactic point sources discarded from the analysis
         because they located are behind H~{\sc ii} regions or supernova
         remnants, detected in total intensity or H$\alpha$. The
         columns give the longitude $l$ and latitude $b$ of the source
         in degrees, its RM in rad~m$^{-2}$ and the reason for
         flagging it: Stokes~I or H$\alpha$. \label{t:flag}}
\end{center}
\end{table*}

\begin{figure}[t]
\centerline{\psfig{figure=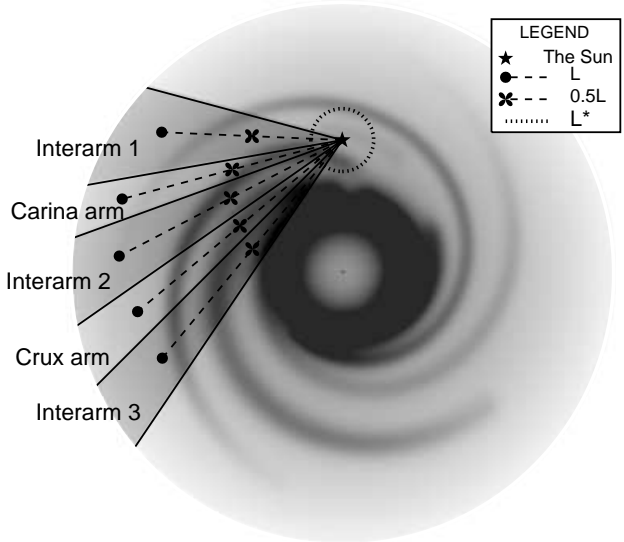,width=.5\textwidth}}
\caption{Bird's-eye view of the Galaxy, with the longitude ranges that
define the 'spiral arms' and 'interarm regions'. The distance $L$
which contains 90\% of the total electron density along the line of
sight is given by the length of the dashed line. The distances $0.5L$
at which the regular magnetic field strength is calculated is given by
the black crosses. The dotted circle denotes the distance $L^*$ used
for calculating angular outer scales.}
\label{f:gal}
\end{figure}

The SGPS data probe the inner Galaxy, which includes a number of
spiral arms. Consequently, these data are well-suited to study
differences in the structure in the ISM in spiral arms and in interarm
regions. We constructed second-order structure functions of RM,
$D_{RM}(\delta\theta) = \left< [\mbox{RM}(\theta) -
\mbox{RM}(\theta+\delta\theta)]^2 \right>_{\theta}$, for different
lines of sight in 'spiral arms', i.e.\ lines of sight primarily
through spiral arms, and 'interarm regions', sight lines mostly
through interarm regions, estimated from the spiral arm positions in
\citet{cl02}. The lines of sight used to separate spiral arms and
interarm regions are shown in Fig.~\ref{f:gal}. The error bars
denote errors propagated from uncertainties in the RM values. For
lines of sight at high longitudes close to the Galactic center it is
not possible to distinguish between `spiral arm lines of sight' and
'interarm lines of sight', because spiral arms start running
perpendicular to the line of sight. Therefore, we do not use data at
$l > 326^{\circ}$ in this analysis.

\begin{figure}[t]
\centerline{\psfig{figure=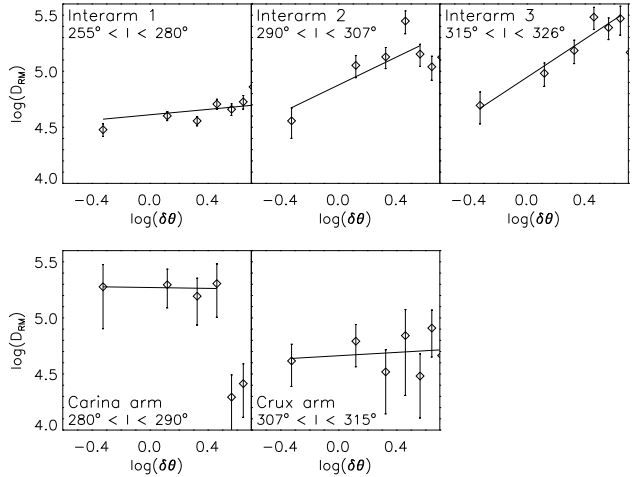,width=.5\textwidth}}
\caption{Structure functions of RM for Galactic interarm regions (top)
         and spiral arms (bottom). The solid lines are linear fits to
         the rising parts of the structure functions.}
\label{f:sf}
\end{figure}

Figure~\ref{f:sf} shows $D_{RM}$ for spiral arms and interarm
regions. The figure is similar to Fig.~1 in \citet{hgm06} except here
we have used the finalized RM list in \citet{bhg07}, and discarded the
27~sources in Table~\ref{t:flag}. The spiral arm structure functions
are flat, while in interarm regions the structure functions rise, and
in two out of three cases show a turnover from a power law at small
scales to flat at the larger scales. The location of the turnover is
interpreted as the largest angular scale of structure in the interarm
regions. Using the argument that the largest angular scales in RM are
probably coming from nearby, assuming a distance of 2~kpc yields an
outer scale at a spatial scale of about $100$~pc. For the spiral arms
we can only say that the outer scale of structure, i.e.\ the smallest
scale we probe, is smaller than about 10~pc \citep{hgb06}.

The amount of sources used to estimate the structure functions are 50,
20 and 18 for the respective interarm regions, and 8 sources for each
of the spiral arms. Given the low source density in the spiral arms,
we assess the reliability of the results here. The bin size is
somewhat restricted due to the paucity of sources, but bin sizes
between 0.5$^{\circ}$ and 1$^{\circ}$ are reasonable and yield
comparable results.

\begin{figure}[t]
\centerline{\psfig{figure=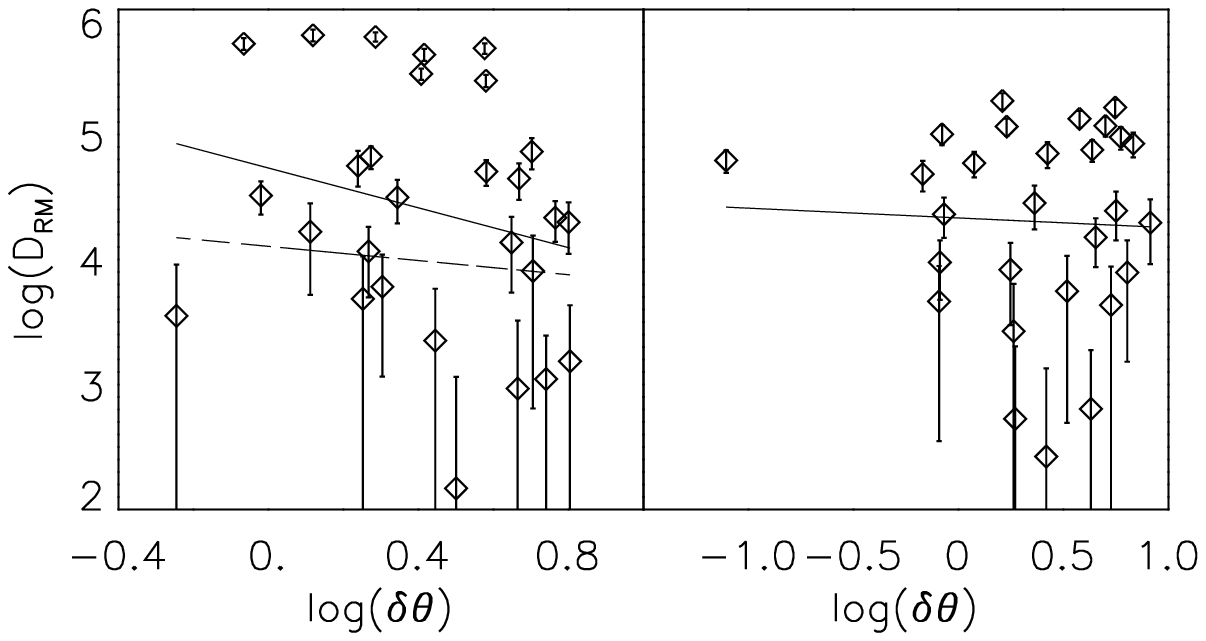,width=.5\textwidth}}
\caption{Structure functions of RM for the Carina spiral arm (left and
the Crux arm (right). Each data point denotes a single source
pair. The solid lines denote linear fits of all data. The dashed line
in the Carina arm plot indicates a linear fit after discarding of the
upper row of points at $\log(D_{RM}) > 5.3$, which are all caused by one source of extreme
RM.}
\label{f:sfarm}
\end{figure}

Figure~\ref{f:sfarm} shows the structure functions of RMs towards the
Carina and Crux arms without any binning of sources. The solid lines
give linear fits of the data, confirming our statement that the
structure function in the spiral arms is flat. In the Carina arm, the
seven uppermost points (all at $\log(D_{RM}) > 5.3$) are the
combination of one source with extreme RM and all other sources in the
region. The presence of this extreme source makes clear why the two
data points on the largest scales are lower than the other points: on
these scales the extreme RM source does not contribute. Furthermore,
this explains why the amplitude of the structure function is higher in
the Carina arm than in the Crux arm: omitting this source yields
comparable amplitudes for both arms.

However, we are hesitant to discard sources on the basis of extreme RM
alone, as there is no reason per se why these sources would not be
part of the spectrum. Therefore, we only omit sources visibly behind a
discrete structure, as discussed above, and leave this extreme RM
source in the dataset, while commenting on changes when this source is
omitted.

We can estimate the outer scale from modeling of the amplitude and
slope of the structure functions, or from the amount of depolarization
of the point sources by the Galaxy. These two methods will be
discussed in Sections~\ref{s:sf} and~\ref{s:depol}, respectively.


\section{Outer scale from rotation measure structure functions}
\label{s:sf}

The structure function slopes in Fig.~\ref{f:sf} are $0.32\pm0.05$,
$1.09\pm0.18$ and $0.71\pm0.17$ for interarm regions 1, 2, and 3,
respectively, much shallower than those expected from a Kolmogorov
structure function, which has a slope $m=5/3$. However, we argue in
this section that on smaller scales the RM structure function has to
turn over to a steeper slope to be consistent with observations of
electron density fluctuations on smaller scales
\citep[e.g.,][]{ars95}; additional arguments for a steeper RM power
spectrum on smaller scales are given in Section~\ref{s:steep}.

Minter \& Spangler (1996, hereafter MS96) developed a formalism to
describe the structure function of RM assuming power spectra in
magnetic field fluctuations and in electron density fluctuations which
are zero-mean, isotropic, Gaussian, and independent.  Assuming
Kolmogorov turbulence the RM structure function, $D_{RM}$, can be
described as:
\begin{eqnarray}
  D_{\mbox{RM}}\!\!\!\!&=&\!\!\!\!\! \left\{\!\right. 251.226\left[\right. \left({\small 
    \frac{n_{e0}}{0.1~\mbox{cm}^{-3}}}\right)^2\! \left({\small 
    \frac{C_B^2}{10^{-13}\mbox{ m}^{-2/3}\mu\mbox{G}^2}}\right) \nonumber\\
    &&+\left({\small \frac{B_{0\parallel}}{\mu\mbox{G}}}\right)^2\! 
    \left({\small\frac{C_n^2}{10^{-3} \mbox{ m}^{-20/3}}}\right)\left.\right] 
     \nonumber\\
    &&+\,23.043\left({\small\frac{C_n^2}{10^{-3} \mbox{ m}^{-20/3}}}\right
)\nonumber\\
    &&\times\left({\small \frac{C_B^2}{10^{-13}\mbox{ m}^{-2/3}\mu\mbox{G}^2}}\right)\!
    \left({\small \frac{l_0^K}{\mbox{pc}}}\right)^{2/3} \!\!\left.\right\} \nonumber\\
    &&\times\left({\small \frac{L}{\mbox{kpc}}}\right)^{8/3}\!
    \left({\small \frac{\delta\theta}{\mbox{deg}}}\right)^{5/3}
    \label{e:ms96}
\end{eqnarray}
where $n_{e0}$ is the mean electron density, $B_{0\parallel}$ is the
mean magnetic field strength along the line of sight, $l_0^K$ the
outer scale of the Kolmogorov turbulence, and $L$ the length of the
line of sight. The coefficients $C_B^2$ and $C_n^2$ are defined in the
description of the magnetic field and density fluctuations as power
laws with the same outer scale $l_0^K$ and spectral index $\alpha$
such that
\begin{equation}
  \left<\delta B_i(\mathbf{r_0})\delta B_i(\mathbf{r_0+r}) \right> =
  \int d^3q \frac{C_B^2 \mbox{e}^{-i\mathbf{q}\cdot\mathbf{r}}}{(q_0^2+q^2)^{\alpha/2}}
\end{equation}
where wave number $q = 2\pi/r$ and $q_0 = 2\pi/l_0^K$. A similar
expression applies for $\left<\delta n(\mathbf{r_0})\delta
n(\mathbf{r_0+r}) \right>$. The spectral index of the power spectrum
is $\alpha=11/3$ for Kolmogorov turbulence, and is related to the
slope of the structure function $D_{RM}\propto r^m$ as $m=\alpha-2$
for $2<\alpha<4$ \citep{r77}.  Equation~\ref{e:ms96} is only valid if
$\delta\theta L/l_0^K<1$, which is only true on scales smaller than
what we observe. However, arguing that the slope turns over to
Kolmogorov on smaller scales, we use a Kolmogorov dependence according
to equation~(\ref{e:ms96}) on scales $l < l_0^K$, which flattens to
the shallow or flat observed spectra for $l>l_0^K$.

Most of the input parameters in equation~(\ref{e:ms96}) are known, so
that by joining this equation with RM structure functions in
Figure~\ref{f:sf} the outer scale of Kolmogorov turbulence $l_0^K$ can
be computed. The amplitude of the electron density fluctuations
$C_n^2$ is taken to be $C_n^2= 10^{-3}$~m$^{-20/3}$ \citep{ars95}, and
the magnetic field fluctuations $C_B^2$ can be derived from the
observed value from MS96 as $C_B^2 = 5.2 B_{ran}^2 (l_0^K)^{-2/3}
\mu$G$^2$~m$^{-2/3}$, where $B_{ran}$ is the strength of the random
component of $B$ in $\mu$G and $l_0^K$ is in parsecs. Values for the
mean electron density $n_{e0}$, mean magnetic field $B_0$, random
magnetic field $B_{ran}$, and distance to the emission will be derived
in the following subsections.

\subsection{Magnetic field}

A number of estimates of the total magnetic field strength in the
Galaxy, based on a number of observations, indicate that the total
magnetic field strength is around $6~\mu$G at the Solar radius.
\citet{h95} gives an extended discussion about the different ways to
determine Galactic magnetic field strengths, i.e.\ using the
synchrotron emissivity under the assumption of minimum energy or
minimum pressure, or with the cosmic ray density measured in the solar
neighborhood, and using Faraday rotation from pulsars. His estimates
for the total magnetic field in the solar neighborhood range from 4~to
$7.4~\mu$G for different assumptions, while he estimates $B_{tot}\sim
7.6 - 11.2~\mu$G at Galactocentric radius $R_{Gal}=4$~kpc. More
recently, \citet{smr00} modeled cosmic ray evolution and propagation
in the Milky Way, using constraints from synchrotron and $\gamma$-ray
emission. Their results indicate a total magnetic field strength of
6.1~$\mu$G at the solar circle, increasing exponentially towards the
inner Galaxy with a scale length of 10~kpc. These measurements are in
good agreement with each other, with results from synchrotron
radiation \citep{bbm96} and from pulsars \citep{hfm04}. 

We can estimate the relative strengths of the magnetic field in the
spiral arms and interarm reasons from Fig.~7b in \citet{bkb85}, which
shows that the mean volume emissivity in the spiral arms is twice as
high as the mean volume emissivity in the interarm regions, i.e.\
$\epsilon_{arms} = 2\epsilon_{interarms} = 1.5\langle\epsilon\rangle$,
where the average volume emissivity is
$\langle\epsilon\rangle=0.5(\epsilon_{arms}+\epsilon_{interarms})$.
Therefore, the total magnetic field strength in the spiral arms must
be $(3/2)^{2/7}$ that in the interarm regions\footnote{We follow the
argument put forth in \citet{h95}, although he uses $\epsilon_{arms}
\approx 2\langle \epsilon \rangle = 2(\epsilon_{arms} +
\epsilon_{interarms})$. His approximation $\epsilon_{interarms}
\approx 0$ is reasonable in the outer Galaxy. However, the
\citet{bkb85} synchrotron maps indicate that $\epsilon_{arms} =
2\epsilon_{interarms}$ is more appropriate in the inner Galaxy.},
where the exponent $2/7$ occurs because in the minimum energy
approximation $\langle B \rangle \propto \epsilon^{2/7}$.
We adopt the dependence of the magnetic field strength on
Galactocentric radius in \citet{smr00}, and correct for the relative
strengths in the arms and interarm regions as given above, leading to
the total field strengths given in Table~\ref{t:out}. The regular
magnetic field component $B_{reg}$ is estimated by \citet{hml06} from
pulsar dispersion and rotation measures as $B_{reg}=2.1\pm0.3~\mu$G,
increasing exponentially inwards with a scale length of 8.5~kpc. We
adopt these values for the regular magnetic field strength, so that
$B_{ran} = \sqrt{B_{tot}^2-B_{reg}^2} \approx 0.94B_{tot}$, independent
of Galactocentric radius. The component of the regular magnetic field
parallel to the line of sight $B_{0\parallel}$ in
equation~(\ref{e:ms96}) is evaluated through $B_{0\parallel} =
B_0\cos\beta$, where $\beta$ is the angle between the line of sight
and the regular magnetic field ${\mathbf B_0}$. A pitch angle of
$-12^{\circ}$ \citep{v04} is assumed, but a pitch angle of 0$^{\circ}$
gives no significant changes.

\subsection{Electron density}

The average electron density used for each line of sight is derived
from the NE2001 electron density model \citep{cl02}. The electron
density was evaluated from the model for a particular line of sight
centrally through each arm and each interarm region. The adopted
average electron density is the average density over the adopted line
of sight. Most of the structure in NE2001 is on such large scales that
a change in direction of the line of sight of a degree or so does not
influence the results significantly in any of the lines of sight. The
adopted values for the mean electron density along each line of sight
can be found in Table~\ref{t:out}.

\subsection{Distances}
\label{s:dist}

Two different distances for modeling are needed: $L$, the total path
length of the Faraday rotating material, and $L^*$, the distance used
to convert the outer scale of fluctuations from angular to spatial
scales, both of which are shown in Fig.~\ref{f:gal}. The distance $L$
is chosen as the distance to the point for which 90\% of the electron
density is contained along the path, and is given in
Table~\ref{t:out}. Subsequently, the distance from the sun at which
$B_{tot}(r)$ is computed for each sight line is $0.5L$.

If the statistical properties of the medium do not change along a
given line of sight, the distance at which a certain angular scale is
expected to correspond to the largest spatial scale is a small
distance. Consequently, $L^*$, which corresponds to the largest
spatial scale $l_0^K$, should be different from $L$, and is estimated
to be 2~kpc.

These distances are very rough estimates. However,we show in
Section~\ref{s:sens} that the sensitivity of our conclusions to the
anticipated error in distances is low.

\subsection{Results from RM structure functions}
\label{s:results}

\begin{figure}[t]
\centerline{\psfig{figure=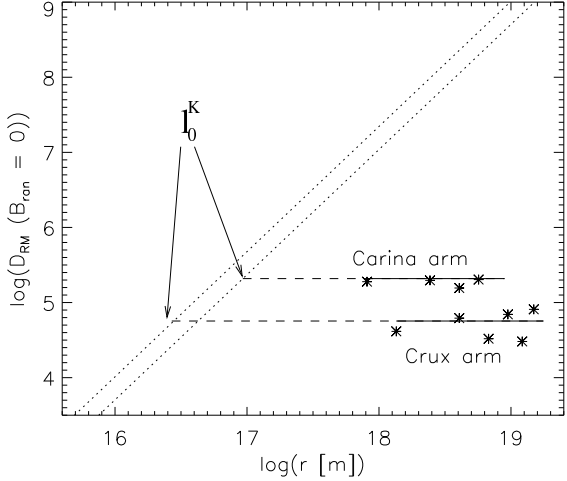,width=.45\textwidth}}
\caption{Estimate of the structure function of RM, $D_{RM}$, for a
         constant magnetic field (i.e.\ $B_{ran}=0$), with the
         extrapolation from electron density fluctuations measured on
         smaller scales for the input parameters of the Carina and
         Crux arms (dotted lines). The asterisks are data from the
         Carina and Crux arms, with solid line fits. The dashed lines
         show the suggested connection to the data and the arrows
         denote the outer scale $l_0^K$.}
\label{f:sfex}
\end{figure}

We extrapolate the observed slopes to smaller scales, with a
steepening to a Kolmogorov spectrum at scale $l_0^K$, which is the
outer scale of the Kolmogorov part of the spectrum.  The constraint of
equal amplitudes between the Kolmogorov structure function on the
small scales and the shallower structure function on the larger scales
at turnover scale $l_0^K$ yields

\begin{equation}
  D_{RM}(\delta\theta) \propto \left\{
  \begin{array}{ll}
     \delta\theta^{5/3}           & \mbox{for}~\delta\theta \le l_0^K/L^* \\
     (l_0^K/L^*)^{5/3-m}\delta\theta^{m}  & \mbox{for}~\delta\theta \ge l_0^K/L^*
  \end{array}
  \right.\label{e:mssf}
\end{equation}

where $m$ is the spectral index of the shallower structure function. In
reality the structure function will not make a sharp break as
described here and instead will show a gradual turnover over a range
of scales \citep[e.g.,][]{hgm04}, but we use this parameterization as
a good first approximation. Combining equations~(\ref{e:ms96})
and~(\ref{e:mssf}) for $\delta\theta \ge l_0^K/L^*$ yields

\begin{eqnarray}
  D_{RM}(\delta\theta) &=& \left[\right. C_1 n_{e0}^2 B_{tot}^2 + C_2 B_{tot}^2(l_0^K)^{2/3} \nonumber \\
    && +C_3 B_{tot} \cos\beta \left.\right]\nonumber \\
    && \times L^{8/3}\left(\frac{l_0^K}{L^*}\right)^{5/3-m}\delta\theta^m,\label{e:ms2}
    \label{e:ms2}
\end{eqnarray}
where $C_1=488.37$, $C_2=0.04$ and $C_3=0.08$ are constants derived
from the known variables in equation~(\ref{e:ms96}), and $m$ is the
spectral index of the structure function. We fit
equation~(\ref{e:ms2}) to the observed structure functions to obtain
estimates for the outer scale of the Kolmogorov part of the power
spectrum $l_0^K$(SF), with results given in Table~\ref{t:out}.

To compare our results with the Kolmogorov spectra found in electron
density, we evaluate the structure function of RM caused by electron
density fluctuations only ($B_{ran}=0$), shown as the dotted lines in
Fig.~\ref{f:sfex} for the input parameters of the Carina and the Crux
arm. This is a lower limit for the structure function of RM if the
power spectrum of electron density on small scales is extrapolated to
the large (parsec) scales discussed here. The spiral arm data, plotted
as asterisks in Fig.~\ref{f:sfex}, fall {\it below} the extrapolation
of $D_{RM}$ to large scales, indicating that $D_{RM}$ has to turn over
at smaller scales where the dotted and dashed lines in
Fig.~\ref{f:sfex} meet. The same argument holds for the interarm data,
which are not shown in the Figure for clarity. Although the input
parameters are uncertain, this result is remarkably stable against
variations of the input parameters (Section~\ref{s:sens}).
For fluctuations in electron density only, this turnover scale is
around 3~pc (1~pc) for the Carina (Crux) arm. If magnetic field
fluctuations are present as well, the turn over scale $l_0^K$ is a
little smaller, as noted in Table~\ref{t:out}. Therefore, the
Kolmogorov spectrum in electron density on small scales, if
extrapolated towards larger scales, does not extend all the way up to
scales of $\sim100$~pc as previously assumed, but displays a break to
a shallower slope (interarms) or a constant value (spiral arms). If it
did, equation~(\ref{e:ms2}) demonstrates that at a scale $r = 100$~pc,
$D_{RM}$ would be 2.3 orders of magnitude higher than the observed
values.

\subsection{Sensitivity of results to input parameters}
\label{s:sens}

Due to the non-straightforward dependence of $D_{RM}$ on $l_0^K$ in
equations~(\ref{e:mssf}) and~(\ref{e:ms2}), we tested numerically how
sensitive the results are to variations in the input parameters $L$,
$L^*$ and $n_e$. If the path length $L$ is increased or decreased by
30\%, this will decrease or increase the resulting outer scale $l_0^K$
by a factor of 50\% . The same effect is seen for an increase or
decrease in $L^*$ or $n_e$ by a factor two. Although it makes sense to
assume that the largest angular scales $\delta\theta_0$ correspond to
the largest spatial scales $l_0^K$ at some nearby position along the
line of sight, as we have done, even if we assume that the largest
spatial scales are at the midway distance along the line of sight,
$l_0^K = 0.5~L \tan(\delta\theta_0)$, the obtained outer scale $l_0^K
\approx 1-5$~pc for both arms and interarm regions. Also, if the
amplitude of electron density $C_n^2$ were a factor~10 different in
our data from the data in \citet{ars95}, the outer scale would change
by less than 50\%. As the results are fairly robust against reasonable
changes in the input parameters, we feel confident in asserting that
the outer scale of Kolmogorov turbulence $l_0^K$ must be on the order
of a few parsecs. We note, however, that due to the assumptions made
and uncertainties in input parameters, a relatively large uncertainty
in the determined outer scale has to be taken into account.

\section{Outer scale from depolarization of point sources}
\label{s:depol}

\begin{figure*}[t]
\centerline{\psfig{figure=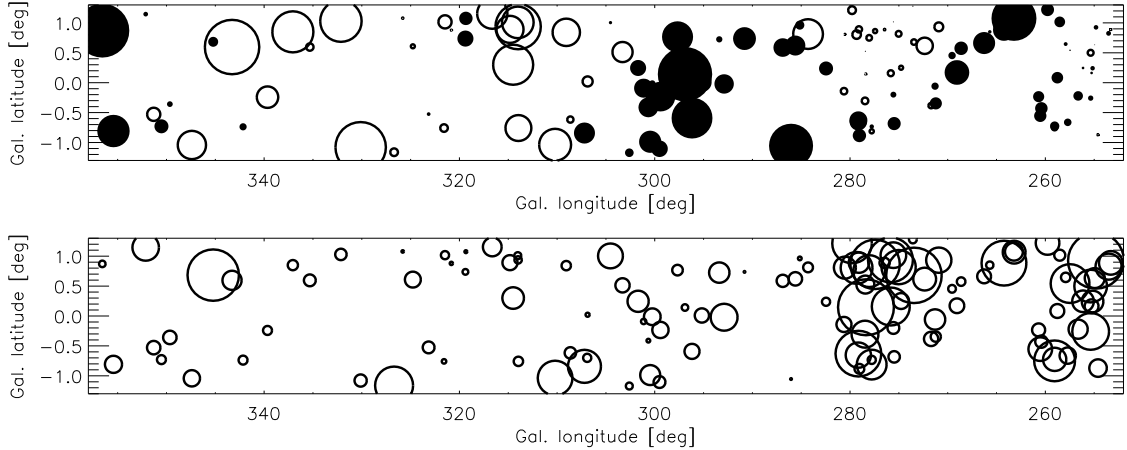,width=.9\textwidth}}
\caption{The SGPS field of view, where circles denote the position of
         extragalactic sources. Top: Filled (open) symbols represent
         positive (negative) RM scaled linearly as $-1000$~rad~m$^{-2}
         <$~RM~$<1000$~rad~m$^{-2}$. Bottom: symbols represent the degree of
         polarization, $p$, scaled between 0.4\% and 13.7\%.}
\label{f:depol}
\end{figure*}

An independent estimate of the outer scales of magneto-ionic structure
can be made from depolarization of extragalactic point sources, if
caused by structure in the foreground ISM on angular scales smaller
than the size of the (unresolved) source.  Intrinsic variations in
polarization angle causing partial depolarization are expected to
arise within any polarized extragalactic source. Indeed, no source in
our sample exhibits the theoretical maximum degree of polarization of
around 70\% \citep{p70} but instead observed degrees of polarization
lie typically below 10\%. Depolarization by foreground components can
be caused by beam depolarization due to magnetic field and/or electron
density fluctuations on scales smaller than the source size \citep[see
e.g.\ ][]{ghs05} or by bandwidth depolarization \citep[see e.g.\
][]{st07}.  Bandwidth depolarization is given by $p = p0
\sin(\Delta\theta)/\Delta\theta$, where $\Delta\theta =
2\mbox{RM}c^2\Delta\nu/\nu^3$ \citep{gw66}. However, significant
bandwidth depolarization across our frequency band $\Delta\nu$
(8~MHz) can only be achieved by $RM \ga 5700$~rad~m$^{-2}$, much
higher than observed RMs. Therefore, any foreground depolarization in
our data is caused by beam depolarization across the face of the
source rather than bandwidth depolarization.

\begin{figure}[t]
\centerline{\psfig{figure=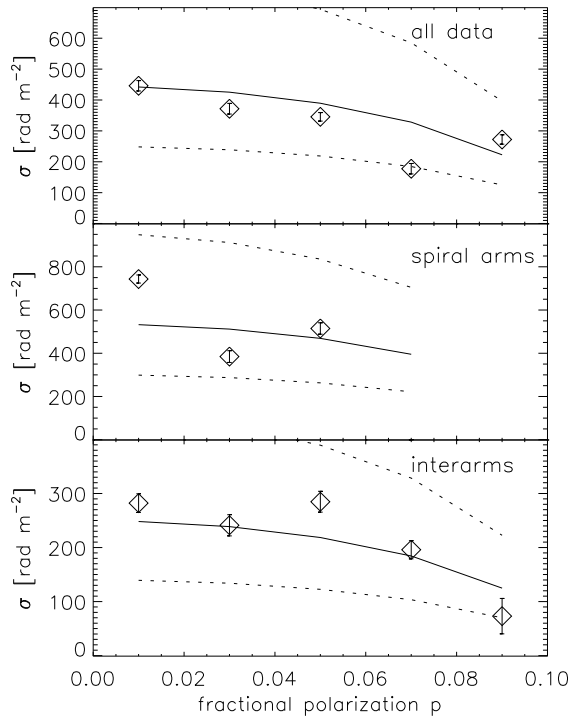,width=.45\textwidth}}
\caption{Diamonds show standard deviation of RM, $\sigma$, versus
         degree of polarization, $p$, for all data (top), spiral arms
         only (middle) and interarm regions only (bottom).
         Sources are binned in degree of polarization with bin width
         $\Delta p = 0.02$. The solid lines are model predictions
         from equation~(\ref{e:tribble}), for $l_0^K =$~7.7~pc, 8.7~pc
         and 3.5~pc for the top, middle, and bottom panels,
         respectively. The dashed lines show the models for $l_0^K$ a
         factor 2 larger and smaller.}
\label{f:rm_p}
\end{figure}

Figure~\ref{f:depol} shows RMs in the upper panel and degree of
polarization, $p$, in the bottom panel. A clear anti-correlation
between $|$RM$|$ and $p$ is visible especially at the lower
longitudes. As the scale of the structure in RM and $p$ is several
degrees, this cannot be intrinsic to the sources but instead must be
caused by the Galactic ISM. This agrees with the results of
\citet{ghs05} who found an anticorrelation between degree of
polarization and H$\alpha$, which is correlated with $|$RM$|$. Due to
the power law nature of RM as a function of scale, a high RM also
indicates large fluctuations in RM. This is shown in
Figure~\ref{f:rm_p}, which shows the standard deviation in RM as a
function of fractional polarization, for all data together and for the
spiral arms and interarm regions separately. As expected, the data in
the spiral arms show a higher standard deviation in RM and a lower
fractional polarization than in the interarm regions.

The Galactic component of depolarization can be estimated from the 
power law spectrum of RM fluctuations. The angular size of the outer
scale of structure $\theta_0^K$ is much larger than the angular source
size $\theta_{src}$. In this approximation $\theta_0^K>>\theta_{src}$,
the depolarization by a power spectrum of RM fluctuations is given by
the degree of polarization $p$, adapted from \citet{t91} as:
\begin{equation}
  \langle (\frac{p(\lambda)}{p_0})^2\rangle \approx 1 - 4 \sigma^2 
  \lambda^4 2^{m/2} \left(\frac{r_{src}}{l_0^K}\right)^m 
  \Gamma(1+\frac{m}{2})
  \label{e:tribble}
\end{equation}
where $p_0$ is the degree of polarization of the extragalactic source
when its radiation arrives at the Milky Way, and $r_{src}$ is the size
of the source. The RM standard deviation $\sigma$, structure function
slope $m$ and outer scale $l_0^K$ are related via the structure
function $ D_{\mbox{RM}}(r) = 2 \sigma^2 (r/l_0^K)^m$ for $r < l_0^K$.

A decrease in degree of polarization $p$ with increasing standard
deviation in RM $\sigma$, as predicted by equation~\ref{e:tribble}, is
visible in Figure~\ref{f:rm_p} for all data together and for the
spiral arms and interarm regions separately. As expected, the data in
the spiral arms show a higher standard deviation in RM and a lower
fractional polarization than in the interarm regions. With an estimate
of $p_0$ and $r_{src}$, we can fit $\sigma$ as a function of $p$ to
the data points, and obtain a best-fit outer scale $l_0^K$.

The amount of intrinsic depolarization resulting in polarization
degree $p_0$ can be estimated from the extragalactic sources observed
around the LMC (Gaensler et al.\ 2005) to be 10.4\%, which we assume
is the average degree of polarization of unresolved point sources at
1.4~GHz for which all depolarization is intrinsic. With these
assumptions, the depolarization below 10.4\% is then due to the
variations in Galactic RM across the face of the source with a median
size of 6~arcsec for this flux density range (Gaensler et al.\
2005). Note that the percentage of intrinsic depolarization is higher
than the actual average degree of polarization due to a selection of
strong, highly-polarized sources over weak, weakly polarized ones, and
due to selection of sources with linear $\phi(\lambda^2)$
behavior. However, since we are interested in the relative
depolarization only, this selection effect does not influence our
conclusions.

In the spiral arms it is straightforward to use
equation~(\ref{e:tribble}) to determine the outer scale $l_0^K$ needed
to obtain the observed depolarization. Assuming Kolmogorov turbulence
($m = 5/3$), we determine the value of the RM standard deviation
$\sigma$ from the structure function saturation level.  For the
interarm regions we observe a spectrum which is considerably shallower
than a Kolmogorov spectrum. Assuming that this spectrum turns over to
a steeper Kolmogorov spectrum towards small scales, the Kolmogorov
slope on small scales will dominate the depolarization of the point
sources. Therefore $\sigma$ can be assumed to be the value of the
structure function at the scale $l_0^K$, which is the outer scale of
the Kolmogorov turbulence, i.e.\ the scale at which the Kolmogorov
slope turns over into a shallower slope.

The data are best represented by equation~(\ref{e:tribble}) for
$l_0^K=0.2^{\circ}, 0.25^{\circ}$, and $0.1^{\circ}$ for all data,
spiral arms and interarm regions, respectively (shown in
Figure~\ref{f:rm_p} as a solid line, with outer scales twice higher
and lower indicated by dashed lines), so that the outer scales of
Kolmogorov turbulence are approximately 8.7~pc in the spiral arms and
3.5~pc in the interarm regions. The data in the spiral arms are fairly
sparse - confirming additional depolarization in the spiral arms - so
that the standard deviations and the fit to the model are uncertain.
However, in the interarm regions the depolarization model is a good
fit to the data, and the probability that the standard deviations are
constant with fractional polarization is $<0.1\%$.

The estimates of $l_0^K$ from depolarization are somewhat larger than
$l_0^K$ from the structure function analysis. However, if we consider
that the errors in the distances are large, in conjunction with
assumptions such as the lack of correlation between magnetic field and
electron density, the difference in estimates of $l_0^K$ from the two
methods is not necessarily significant. Certainly both methods
indicate that the outer scale of the Kolmogorov spectrum is likely a
few parsecs, much smaller than the previously assumed value of $\sim
100$~pc.

\section{Expected steep spectrum on small scales}
\label{s:steep}

As argued in the previous section, the RM structure function has to
turn over to steeper slopes on smaller scales to be consistent with
the electron density fluctuation data on smaller scales. There are two
additional reasons why the RM structure functions cannot continue to
have the same shallow or flat slope on smaller scales.

Firstly, a steep Kolmogorov-like magnetic field power spectrum is
indicated by cosmic ray data. As cosmic rays are most effectively
scattered by magnetic field fluctuations on the same scale as their
ion gyro radius, cosmic ray losses as a function of energy are closely
related to the magnetic field power spectrum. The cosmic ray
distribution as calculated from a leaky box model can explain cosmic
ray observational data if the Galactic magnetic field has a power
spectrum with a Kolmogorov spectral index \citep{j88}. Furthermore,
the cosmic ray power spectrum is remarkably smooth\footnote{The
steepening of the spectrum around $3\times10^{15}$~eV (the ``knee'')
and flattening around $3\times10^{18}$~eV (the ``ankle'') are very
slight and not relevant for this argument.} on scales of $10^9$ to
$10^{18}$~eV, corresponding to gyro radii of $10^{20}$ to
$10^{12}$~cm.  Therefore, the magnetic field power spectrum is also
expected to be smooth on these scales.

However, recent numerical simulations show that the magnetic field
power spectrum does not necessarily follow the electron density
spectrum. In fact, magnetohydrodynamic simulations in the limit of a
weak homogeneous magnetic field show that the magnetic field
fluctuations are all concentrated on scales much smaller than those
under discussion here \citep{scm02}. This would indicate a flat
structure function of magnetic field on larger scales. At first sight,
this theory agrees with the observations shown in Fig.~\ref{f:sf},
where the weak-mean-field approximation could be applicable to the
spiral arms, and structure functions would be expected to be
flat. However, this would indicate that magnetic field fluctuations of
a few microgauss would be present on scales as small as a fraction of
a parsec.  In this case the degree of polarization, $p$, as a function
of the intrinsic degree of polarization, $p_0$, is given by
\begin{equation}
p = p_0  \mbox{e}^{-2\sigma^2\lambda^4} \approx \mbox{e}^{-309}
\end{equation}
\citep{b66,sbs98} in the approximation that the scale of the
fluctuations is much smaller than the telescope beam. The standard
deviation of RM, $\sigma$, is derived from the flat structure
functions in the spiral arms. Therefore, magnetic field fluctuations
on these scales would completely depolarize the synchrotron radiation
in the Galaxy at 1.4~GHz.  Since we observe polarized radiation at
this frequency coming from all over the Galactic plane
\citep[e.g.][]{tgp03,hgm06}, magnetic field fluctuations cannot remain
at this magnitude towards smaller scales. Instead, we showed in
Section~\ref{s:depol} that the observed amount of depolarization is
consistent with a RM structure function with a Kolmogorov slope and an
outer scale of a few parsecs or smaller.

\begin{table*}[!t]
\begin{center}
\begin{tabular}{ccccccccc}
region   & longitude&$B_{tot}$& $n_{e0}$ & $L$ &$L^*$&$l_0^K$(SF)&$l_0^K$(depol) \\
         & [$^{\circ}]$ &[$\mu$G]&[cm$^{-3}$]&[kpc]&[kpc]& [pc]  & [pc]\\
\hline
Interarm 1 & 255 - 280  & 4.5 & 0.03  & 11.5 & 2  & 2.3  & 3.5\\
Interarm 2 & 290 - 305  & 4.6 & 0.045 & 16   & 2  & 0.8  & 8.7\\ 
Interarm 3 & 315 - 326  & 5.2 & 0.075 & 18   & 2  & 0.3  & 3.5\\
Carina arm & 280 - 290  & 8.3 & 0.06  & 14.5 & 2  & 2.4$^*$& 8.7\\
Crux arm   & 305 - 315  & 9.0 & 0.07  & 17   & 2  & 1.0  & 3.5\\
\hline
\multicolumn{9}{l}{$^*$ This value is 0.65~pc if the extreme RM source discussed in Section~\ref{s:data} is omitted.}
\end{tabular}
\caption{ISM parameters for three interarm regions and the Carina and
         Crux spiral arms. The parameter $n_{e0}$ is the electron
         density averaged over the line of sight; $L$ and $L^*$ are
         lines of sight as described in the text; and $B_{tot}$ is the
         total magnetic field strength. The outer scale of the
         Kolmogorov power spectrum $l_0^K$ is given as obtained from
         structure functions (SF) and from depolarization of point
         sources (depol).\label{t:out}}
\end{center}
\end{table*}

\section{Discussion}
\label{s:disc}

The conclusion that the outer scale of turbulence in the spiral arms
is observed to be on the order of a few parsecs is not expected or
straightforward. Based on evidence discussed in the introduction,
outer scales of turbulence are expected to be on the order of 100~pc,
which is one or two orders of magnitude higher than those observed
here.  Earlier work noted the small outer scale of fluctuations in the
spiral arms, but attributed that outer scale likely to \ion{H}{2}
regions along the line of sight \citep{hgb06}. \ion{H}{2} regions are
ubiquitous enough in the spiral arms to dominate fluctuations in RM,
and this solution would reconcile a larger outer scale of turbulence
with a smaller observed outer scale of RM fluctuations. However, if
this were the case, the amplitude of the RM structure functions would
lie above the amplitude of the turbulence structure function. Our
estimate of the amplitude of the RM structure functions indicates that
the observed structure functions lie {\it far below} the lower limit
for the RM structure function if extrapolated from electron density
fluctuations on small scales. If the fluctuations in RM associated
with Kolmogorov turbulence were to continue up to scales of a hundred
parsecs, the observed amplitude of the structure function of RM would
be much higher than that observed.

In the interarm regions, a plausible option is multiple scales of
energy input: for supernova-driven turbulence, the outer scale is
believed to be about 100~pc (as observed). However, if energy sources
such as stellar winds or outflows, interstellar shocks or H~{\sc ii}
regions input a significant amount of energy into the interstellar
turbulence on smaller scales (typically parsecs, \citet{mk04}), this
may flatten the structure function on scales of $\sim 1$~pc to scales
of $\sim 100$~pc, consistent with our observations.

This does not contradict earlier studies that reported outer scales of
the order of 100~pc. \citet{sc86} found larger outer scales, but their
data included large parts of the sky, most or all of which were
located at higher latitudes. \citet{ccs92} present some data in the
Galactic plane, but the outer scale of fluctuations in those data is
not well determined due to the paucity of data and the included
geometrical component of the magnetic field. Leahy's (1987) structure
functions of RMs of sources in the Galactic plane are consistent with
an outer scale of a parsec, as is the analysis of DM variations of
close pulsar pairs in globular clusters \citep{ss02,r07}.

So the picture arises of a smaller energy input scale of turbulence or
fluctuations in the magneto-ionized ISM in the Galactic plane, while
larger-scale structure exists in the Galactic thick disk or
halo. Indications of increasing correlation lengths of the magnetic
field with height above the galaxy case have been found in a number of
external galaxies \citep{dkw95}.

\section{Summary and conclusions}
\label{s:sum}

Faraday rotation measurements of polarized extragalactic sources
behind the inner Galactic plane have been used to study the
characteristics of the magnetized, ionized interstellar medium in the
plane, in particular in the spiral arms and in interarm
regions. Rotation measure structure functions show a shallow slope in
the interarm regions and saturation on a scale of $\sim 100$~pc, i.e.\
there are no fluctuations on scales larger than the saturation
scale. Flat structure functions in the spiral arms indicate that the
outer scale of RM fluctuations in the spiral arms is smaller than
$\sim 10$~pc, the smallest scale observed. These shallow and flat
structure functions must turn over to steeper slopes towards smaller
scales for three reasons: (1) to match up with the electron density
power spectrum on subparsec scales, assuming the large and small scale
datasets are part of the same power spectrum; (2) a shallow RM
structure function on smaller scales would give more depolarization
than observed; and (3) cosmic ray distribution data and the smooth
cosmic ray power spectrum indicate a smooth magnetic field power
spectrum with a slope similar to the Kolmogorov slope. The scale of
the break in the structure function is the outer scale of the
Kolmogorov power spectrum $l_0^K$, and is estimated using two
independent methods: the analysis of RM structure functions, and by
modeling the depolarization of the extragalactic sources. Given the
large uncertainties in input parameters, both methods agree reasonably
well and imply an outer scale of the Kolmogorov slope of a few
parsecs. This estimate is almost two orders of magnitude smaller than
the generally assumed outer scale of ISM turbulence of $\sim
100$~pc. However, extrapolating the observed electron density
fluctuations on small scales to parsec scales shows that the amplitude
of the structure function would be orders of magnitude higher if the
Kolmogorov spectrum did in fact extend out to 100~pc. Instead, the
outer scale of Kolmogorov turbulence $l_0^K$ that we obtained from our
observations indicates that energy in ISM turbulence is injected on
scales of a parsec rather than 100~pc. This is the main energy
injection scale in the spiral arms, which show flat structure
functions on scales larger than that. In the interarm regions, the
structure functions keep rising although not as steep as Kolmogorov
turbulence, indicating an additional source of structure. We propose
that in the spiral arms stellar energy sources such as stellar winds
and protostellar outflows are the predominant sources of turbulence,
whereas in the interarm regions there is evidence of energy injection
on larger scales, most likely caused by supernova remnant and
superbubble expansion.

\acknowledgments

The authors thank Anne Green and John Dickey for helpful comments to
the manuscript, and Katia Ferri\`ere for enlightening discussions. The
ATCA is part of the Australia Telescope, which is funded by the
Commonwealth of Australia for operation as a National Facility managed
by CSIRO.  The National Radio Astronomy Observatory is a facility of
the National Science Foundation operated under cooperative agreement
by Associated Universities, Inc.

{\it Facilities:} \facility{Compact Array}.


\begin{thebibliography}{}
\bibitem[Armstrong et al.(1995)]{ars95} 
  Armstrong, J.~W., Rickett, B.~J., \& Spangler, S.~R. 1995, ApJ,
  443, 209 
\bibitem[Balbus \& Hawley(1991)]{bh91}
  Balbus, S. A., \& Hawley, J. F. 1991, ApJ, 376, 214
\bibitem[Beck et al.(1996)]{bbm96} 
  Beck, R., Brandenburg, A., Moss, D., et al. 1996, ARA\&A, 34, 155 
\bibitem[Beuermann et al.(1985)]{bkb85}
  Beuermann, K., Kanbach, G., \& Berkhuijsen, E.~M. 1985, A\&A, 153, 17
\bibitem[Brown et al.(2007)]{bhg07} 
  Brown, J.~C., Haverkorn, M., Gaensler, B.~M., Taylor, A.~R.,
  Bizunok, N.~S., McClure-Griffiths, N.~M., Dickey, J.~M., \& Green,
  A.~J. 2007, ApJ, 663, 258
\bibitem[Burn(1966)]{b66} 
  Burn, B. J. 1966, MNRAS, 133, 67
\bibitem[Clegg et al.(1992)]{ccs92} 
  Clegg, A. W., Cordes, J. M., Simonetti, J. M., \& Kulkarni,
  S. R. 1992, ApJ, 386, 143
\bibitem[Cordes \& Lazio(2002)]{cl02} 
  Cordes, J.~M., \& Lazio, T.~J.~W. 2002, preprint (astro-ph/0207156)
\bibitem[Dumke et al.(1995)]{dkw95} 
  Dumke, M., Krause, M., Wielebinski, R., \& Klein, U. 1995, A\&A,
  302, 691
\bibitem[Elmegreen et al.(2003)]{eel03} 
  Elmegreen, B.~G., Elmegreen, D.~M., \& Leitner, S.~N. 2003, ApJ,
  590, 271
\bibitem[Elmegreen et al.(2001)]{eks01} 
  Elmegreen, B.~G., Kim, S., \& Staveley-Smith, L. 2001, ApJ, 548, 749
\bibitem[Finkbeiner(2003)]{f03}
  Finkbeiner, D.~P. 2003, ApJS, 146, 407
\bibitem[Gaensler et al.(2005)]{ghs05}
  Gaensler, B.~M., Haverkorn, M., Staveley-Smith, L., Dickey, J.~M.,
  McClure-Griffiths, N.~M., Dickel, J.~R., Wolleben, M., 2005,
  Science, 307, 1610
\bibitem[Gardner \& Whiteoak(1966)]{gw66} 
  Gardner, F. F., \& Whiteoak, J. B. 1966, ARAA, 4, 245 
\bibitem[Han et al.(2006)]{hml06}
  Han, J.~L., Manchester, R.~N., Lyne, A.~G., Qiao, G.~J., \& Van
  Straten, W. 2006, ApJ, 642, 868
\bibitem[Han et al.(2004)]{hfm04}
 Han, J.~L., Ferri\`ere, K., \& Manchester, R.~N. 2004, ApJ, 610, 820
\bibitem[Haverkorn et al.(2006a)]{hgb06}
  Haverkorn, M., Gaensler, B.~M.,  Brown, J. C., Bizunok, N. S.,
  McClure-Griffiths, N.~M., Dickey, J.~M., \& Green, A. J. 2006a,
  ApJL, 637, 33
\bibitem[Haverkorn et al.(2006)]{hgm06}
  Haverkorn, M., Gaensler, B.~M., McClure-Griffiths, N.~M., Dickey,
  J.~M., \& Green, A.~J. 2006b, ApJS, 167, 230
\bibitem[Haverkorn et al.(2004)]{hgm04}
  Haverkorn, M., Gaensler, B.~M., McClure-Griffiths, N.~M., Dickey,
  J.~M., \& Green, A. J. 2004, ApJ, 609, 776
\bibitem[Haverkorn et al.(2003)]{hkb03} 
  Haverkorn, M., Katgert, P., de Bruyn, A. G. 2003, A\&A, 403, 1045
\bibitem[Hawley \& Balbus(1991)]{hb91}
  Hawley, J. F., \& Balbus, S. A. 1991, ApJ, 376, 223
\bibitem[Heiles(1995)]{h95} 
  Heiles, C. 1995, in {\it The Physics of the Interstellar Medium and
  Intergalactic Medium}, ed.\ by A. Ferrara, C. F. McKee, C. Heiles,
  \& P. R. Shapiro, p.\ 507
\bibitem[Jokipii(1988)]{j88}
  Jokipii, J. R. 1988, in proceedings of the AIP Conference Radio wave
  scattering in the interstellar medium, ed.\ J.~M. Cordes,
  B.~J. Rickett \& D.~G. Backer, p.\ 48
\bibitem[Kolmogorov(1941)]{k41} 
  Kolmogorov, A.~N., 1941, Dokl.\ Akad.\ Nauk SSSR, 30, 301
\bibitem[Landecker et al.(2006)]{lrw06}
  Landecker, T.~L., Reid, R.~I., Wolleben, M., Reich, W., Kothes, R.,
  Del Rizzo, D., Uyan\i ker, B., Gray, A.~D., \& Taylor, A.~R. 2006,
  AAS, 208, 4909
\bibitem[Lazaryan \& Shutenkov(1990)]{ls90}
  Lazaryan, A. L., \& Shutenkov, V. P. 1990, SvAL, 16, 297L
\bibitem[Leahy(1987)]{l87} 
  Leahy, J. P. 1987, MNRAS, 226, 433
\bibitem[L\"ohmer et al.(2001)]{lkm01}
  L\"ohmer, O., Kramer, M., Mitra, D., Lorimer, D.~R., \& Lyne,
  A.~G. 2001, ApJ, 562, L157
\bibitem[Mac Low \& Klessen(2004)]{mk04}
  Mac Low, M.-M., \& Klessen, R.~S. 2004, Rev.\ Mod.\ Phys., 76, 1, 125
\bibitem[McClure-Griffiths et al.(2005)]{mdg05}
  McClure-Griffiths, N.~M., Dickey, J.~M., Gaensler, B.~M., Green,
  A.~J., Haverkorn, M., Strasser, S. 2005, ApJS, 158, 178
\bibitem[Minter \& Spangler(1996)]{ms96} 
  Minter, A. H., \& Spangler, S. R. 1996, ApJ, 458, 194
\bibitem[Mitra et al.(2003)]{mwk03}
  Mitra, D., Wielebinski, R., Kramer, M., \& Jessner, A. 2003, A\&A,
  398, 993
\bibitem[Norman \& Ferrara(1996)]{nf96} 
  Norman, C.~A., \& Ferrara, A. 1996, ApJ, 467, 280 
\bibitem[Pacholczyk(1970)]{p70}
  Pacholczyk, A.~B. 1970, ``Radio Astrophysics'', W.~H. Freeman and Co.
\bibitem[Ransom(2007)]{r07}
  Ransom, S. M. 2007, in SINS - Small Ionized and Neutral Structures
  in the Diffuse Interstellar Medium, eds.\ M. Haverkorn and
  W.~M. Goss, San Francisco: Astronomical Society of the Pacific,
  p.\ 265
\bibitem[Rickett(1977)]{r77}
  Rickett, B. J. 1977, ARA\&A, 15, 479
\bibitem[Schekochihin et al.(2002)]{scm02} Schekochihin, A., Cowley,
  S., Maron, J., \& Malyshkin, L. 2002, PhRvE, 65, 016305
\bibitem[Selwood \& Balbus(1999)]{sb99}
  Sellwood, J.~A., \& Balbus, S.~A. 1999, ApJ, 511, 660
\bibitem[Shishov et al.(2003)]{sss03}
  Shishov, V.~I., Smirnova, T.~V., Sieber, W., Malofeev, V.~M.,
  Potapov, V.~A., Stinebring, D., Kramer, M., Jessner, A., \&
  Wielebinski, R. 2003, A\&A, 404, 557
\bibitem[Simonetti \& Cordes(1986)]{sc86} 
  Simonetti, J. H., \& Cordes, J. M. 1986, ApJ, 310, 160
\bibitem[Smirnova \& Shishov(2002)]{ss02}
  Smirnova, T. V., \& Shishov, V. I. 2002, Astron. Astrophys.
  Transac., 21, 45
\bibitem[Sokoloff et al.(1998)]{sbs98} 
  Sokoloff, D.~D., Bykov, A.~A., Shukurov, A., Berkhuijsen, E.~M.,
  Beck, R., \& Poezd, A.~D. 1998, MNRAS, 299, 189
\bibitem[Spangler \& Gwinn(1990)]{sg90}
  Spangler, S.~R., \& Gwinn, C.~R. 1990, ApJ, 353L, 29	
\bibitem[Stanimirovi\'c et al.(1999)]{ssd99}
  Stanimirovi\'c, S., Staveley-Smith, L., Dickey, J.~M., Sault, R.~J.,
  Snowden, S.~L. 1999, \mnras, 302, 417
\bibitem[Stil \& Taylor(2007)]{st07}
  Stil, J. M., \& Taylor, A. R., 2007, ApJL, 663, 21
\bibitem[Stinebring et al.(2000)]{ssh00}
  Stinebring, D.~R., Smirnova, T.~V., Hankins, T.~H., Hovis, J.~S.,
  Kaspi, V.~M., Kempner, J.~C., Myers, E., \& Nice, D.~J. 2000, ApJ,
  539, 300
\bibitem[Strong et al.(2000)]{smr00}
  Strong, A.~W., Moskalenko, I.~V., \& Reimer, O. 2000, ApJ, 537, 763
\bibitem[Sun \& Han(2004)]{sh04} 
  Sun, X.~H, \& Han, J.~L. 2004, in The Magnetized Interstellar
  Medium, ed.\ B.~Uyan\i ker, W.~Reich, R.~Wielebinski
  (Katlenburg-Lindau: Copernicus GmbH), 25
\bibitem[Taylor et al.(2003)]{tgp03}
  Taylor, A.~R., Gibson, S.~J., Peracaula, M., Martin, P.~G.,
  Landecker, T.~L., Brunt, C.~M., Dewdney, P.~E., Dougherty, S.~M.,
  Gray, A.~D., Higgs, L.~A., Kerton, C.~R., Knee, L.~B.~G., Kothes,
  R., Purton, C.~R., Uyan\i ker, B., Wallace, B.~J., Willis, A.~G., \&
  Durand, D. 2003, AJ, 125, 3145
\bibitem[Tribble(1991)]{t91}
  Tribble, P.~C. 1991, MNRAS, 250, 726
\bibitem[Vall\'ee(2004)]{v04} 
  Vall\'ee, J.~P. 2004, NewAR, 48, 763
\bibitem[Wang et al.(2005)]{wmj05}
  Wang, N., Manchester, R.~N., Johnston, S., Rickett, B., Zhang, J.,
  Yusup, A., \& Chen, M. 2005, MNRAS, 358, 270
\bibitem[Westpfahl et al.(1999)]{wca99}
  Westpfahl, D.~J., Coleman, P.~H., Alexander, J., \& Tongue, T. 1999,
  AJ, 117, 868
\bibitem[Wolleben et al.(2006)]{wlr06}
  Wolleben, M., Landecker, T.~L., Reich, W., \& Wielebinski, R. 2006,
  A\&A, 448, 411
\bibitem[You et al.(2007)]{yhc07}
  You, X.~P., Hobbs, G.~B., Coles, W.~A., Manchester, R.~N., \& Han,
  J.~L. 2007, astro-ph/0709.0135

\end{thebibliography}
\end{document}